# Robust oblique Target-rotation for small samples

André Beauducel* & Norbert Hilger

*University of Bonn, Germany*

**Abstract**

Introduction: Oblique Target-rotation in the context of exploratory factor analysis is a relevant method for the investigation of the oblique independent clusters model. It was argued that minimizing single cross-loadings by means of target rotation may lead to large effects of sampling error on the target rotated factor solutions.

Method: In order to minimize effects of sampling error on results of Target-rotation we propose to compute the mean cross-loadings for each block of salient loadings of the independent clusters model and to perform target rotation for the block-wise mean cross-loadings. The resulting transformation-matrix is than applied to the complete unrotated loading matrix in order to produce mean Target-rotated factors.

Results: A simulation study based on correlated independent factor models revealed that mean oblique Target-rotation resulted in smaller negative bias of factor inter-correlations than conventional Target-rotation based on single loadings, especially when sample size was small and when the number of factors was large. An empirical example revealed that the similarity of Target-rotated factors computed for small subsamples with Target-rotated factors of the total sample was more pronounced for mean Target-rotation than for conventional Target-rotation.

Discussion: Mean Target-rotation can be recommended in the context of oblique independent factor models, especially for small samples. An R-script and an SPSS-script for this form of Target-rotation are provided in the Appendix.

* Corresponding author, address for correspondence: University of Bonn, Department of Psychology, Kaiser-Karl-Ring 9, 53111 Bonn Phone: +49228 734151, Email: beauducel@uni-bonn.de

Exploratory factor analysis (EFA) is a widely used multivariate method, especially in the context of the development of instruments for psychological assessment. Although confirmatory factor analysis may be used for similar purposes, there is still room for EFA because the expectation of perfect simple structure with one large salient loading of each observed variable on one factor and zero cross-loadings, i.e., an independent clusters model (ICM), may lead to unrealistic simplifications in the context of confirmatory factor analysis. The ICM may cause model misfit in confirmatory factor analysis resulting in model modifications and capitalization on chance (MacCallum et al. 1992). This problem does not occur with the ICM in the context of EFA because Target-rotation towards an ICM in the context of EFA or exploratory structural equation modeling (ESEM; Asparouhov and Muthén 2009) will only provide an orientation of the factor axes so that cross-loadings might be minimized without any consequence for model fit.

The advantage of using Target-rotation in the context of EFA instead of an ICM in the context of confirmatory factor analysis has been demonstrated for the five-factor model of personality (McCrae et al. 1996). Empirical research has also shown that the use of Target-rotation in the context of ESEM allows to avoid an over-estimation of factor inter-correlations that may occur when the ICM is specified in the context of confirmatory factor analysis (Joshanloo 2016). The relationship between cross-loadings, factor inter-correlations, and different criteria of factor rotation has also been investigated in the context of simulation studies (Sass and Schmitt 2010; Schmitt and Sass 2011). Sass and Schmitt (2010) found that the criteria of factor rotation differ in allowing for larger cross-loadings and in the size of the resulting factor inter-correlations.

The relationship between the loading pattern and the factor inter-correlations has also been addressed by Zhang et al. (2019), who extended partial Target-rotation in order to allow for the specification of a Target-matrix for the factor inter-correlations in addition to the Target-matrix for the loadings. With their extension Target-rotation allows for the investigation of hypotheses on the size of factor inter-correlations. Their approach is based on oblique partial Target-rotation (Browne 1972) and the gradient projection algorithm (Bernaards and Jennrich 2005; Jennrich 2002). Moreover, Hurley and Cattell (1962) initially introduced complete oblique Target-rotation providing rotated loadings and estimates for factor inter-correlations when all values of the Target-matrix of loadings are specified.

While Target-rotation allows for a specification of the ICM in the Target-loadings, Target-rotation will typically be performed in order to minimize cross-loadings. Unless specific Target-values are specified for the correlations by means of extended Target-rotation, Target-rotation will modify the factor inter-correlations in order to reduce cross-loadings. If the ICM holds in the population, sampling error will nevertheless lead to some cross-loadings. When the distribution of cross-loadings resulting from sampling error is not perfectly symmetric, minimizing these cross-loadings may affect the factor inter-correlations. Thereby, sampling error may affect factor inter-correlations resulting from Target-rotation. Moreover, when an ICM holds and when single cross-loadings are minimized by Target-rotation, random differences between single cross-loadings may also affect the rotated loading pattern.

It is therefore proposed to minimize the effect of sampling error on the loading pattern and factor inter-correlations resulting from oblique Target-rotation by means of minimizing mean cross-loadings instead of the single cross-loadings. It is expected that using the mean cross-loadings instead of the single cross-loadings for rotation will reduce the effect of sampling error on the results of Target-rotation. The method is termed oblique Mean-Target-rotation (OMT) and may also be of interest when a few substantial cross-loadings occur in the population because it avoids minimizing the single cross-loadings. Thereby, OMT could be helpful for the investigation of departures from the ICM.

After some definitions, the OMT-rotation and a population example will be presented. A simulation study was performed for the oblique ICM to compare OMT-rotation with conventional oblique Target-rotation (OT). Moreover, OMT- and OT-rotation were compared by means of an empirical example. Finally, recommendations for analyses of oblique ICM by means of Target-rotations are discussed.

## Definitions

According to the population common factor model a random vector **x** of $p$ observed variables is explained by a random vector $\xi$ of $q$ common factors and a random vector $\delta$ of $p$ unique factors. This can be written as

$$\mathbf{x} = \mathbf{\Lambda}\boldsymbol{\xi} + \boldsymbol{\delta}, \tag{1}$$

where $\mathbf{\Lambda}$ is the $p \times q$ matrix of factor loadings and $E(\boldsymbol{\xi}\boldsymbol{\xi}') = \mathbf{\Phi}$, $diag(\mathbf{\Phi}) = \mathbf{I}$, $E(\boldsymbol{\delta}\boldsymbol{\delta}') = \mathbf{\Psi}^2 = diag(\mathbf{\Psi}^2)$, and $E(\boldsymbol{\xi}\boldsymbol{\delta}) = \mathbf{0}$. This implies

$$E(\mathbf{xx}') = \mathbf{\Sigma} = \mathbf{\Lambda}\mathbf{\Phi}\mathbf{\Lambda}' + \mathbf{\Psi}^2 = \mathbf{\Lambda}_u\mathbf{\Lambda}_u' + \mathbf{\Psi}^2, \tag{2}$$

where $\mathbf{\Lambda}_u$ is the matrix of common factor loadings for uncorrelated factors, i.e., for $\mathbf{\Phi} = \mathbf{I}$. Oblique target-rotations (Hurley & Cattell, 1962; Browne, 1972) start from an orthogonal loading matrix $\mathbf{\Lambda}_u$, which is mostly the unrotated loading matrix resulting from factor extraction.

**Oblique Mean-Target-rotation**

OMT-rotation starts with an orthogonal Target-rotation (Schönemann 1966) of the unrotated loadings $\mathbf{\Lambda}_u$ towards a loading Target-matrix $\mathbf{\Lambda}_T$ corresponding to a perfect ICM, with

$$\mathbf{\Lambda}_T = (\mathbf{I}_q \otimes \mathbf{1}_{p/q}), \tag{3}$$

where $\mathbf{I}_q$ is a $q \times q$ identity matrix, $\mathbf{1}_{p/q}$ is a $p/q \times 1$ unit-vector representing the Target-loadings and "$\otimes$" denotes the Kronecker-product. The resulting $\mathbf{\Lambda}_1$ represents an orthogonal loading matrix where the salient loadings are a least square approximation of $\mathbf{\Lambda}_T$. Weighted mean loadings are computed for each block of salient loadings

$$\mathbf{\Lambda}_{1m} = (\mathbf{\Lambda}_1 \cdot \mathbf{\Lambda}_T)'\mathbf{\Lambda}_1((\mathbf{\Lambda}_1 \cdot \mathbf{\Lambda}_T)'\mathbf{\Lambda}_T)^{-1}, \tag{4}$$

where " $\cdot$ " is the Hadamard-product. Therefore, $\mathbf{\Lambda}_1 \cdot \mathbf{\Lambda}_T$ yields the weights of the salient loadings so that the cross-loadings are weighted by the salient-loadings of the respective variable on the respective factor. The resulting weighted mean loading matrix $\mathbf{\Lambda}_{1m}$ is a $q \times q$ matrix so that a $q \times q$ identity matrix $\mathbf{I}_q$ can be used as Target-matrix for oblique Target-rotation according to Hurley and Cattell (1962), where the transformation matrix

$$\mathbf{T} = (\mathbf{\Lambda}_{1m}'\mathbf{\Lambda}_{1m})^{-1}\mathbf{\Lambda}_{1m}\mathbf{I}_q, \tag{5}$$

is normalized in order to get

$$\mathbf{T}_n = diag(\mathbf{T}'\mathbf{T})^{-0.5}\mathbf{T}. \tag{6}$$

This transformation matrix is then used for rotation of the complete loadings, with the reference structure

$$\mathbf{\Lambda}_2 = \mathbf{\Lambda}_1\mathbf{T}_n, \tag{7}$$

and the OMT-rotated loading pattern

$$\mathbf{\Lambda}_O = \mathbf{\Lambda}_2 \, diag((\mathbf{T}_n'\mathbf{T}_n)^{-1})^{0.5}, \qquad (8)$$

and the OMT-rotated factor inter-correlations

$$\mathbf{\Phi}_O = (\mathbf{\Lambda}_O'\mathbf{\Lambda}_O)^{-1}\mathbf{\Lambda}_O'(\mathbf{\Lambda}_u\mathbf{\Lambda}_u')\mathbf{\Lambda}_O(\mathbf{\Lambda}_O'\mathbf{\Lambda}_O)^{-1}. \qquad (9)$$

In order to evaluate whether $\mathbf{\Lambda}_{1m}'\mathbf{\Lambda}_{1m}$ is ill-conditioned, the condition-number $\kappa$ is computed (Moler 2008). If $\kappa$ is large, the inversion of the matrix may lead to numerical imprecision. As in ridge regression, there is the option to add small ridge constants when $\kappa$ is large and to retain the solution with the largest mean congruence (Tucker 1951) of $\mathbf{\Lambda}_O$ with $\mathbf{\Lambda}_T$. For large sample sizes and large salient loadings, this option might be irrelevant, but in general, this option could not be harmful as the solution with the best congruence with $\mathbf{\Lambda}_T$ is retained. The loop for the ridge constant can be found in the R- and SPSS-script in Appendix B.

**Population example**

An R-script as well as an SPSS-script based on the example presented here, allowing for OMT and OT-rotation is given in Appendix B. Users of the script may install R-4.3.1 and replace the initial orthogonal loadings by orthogonal loadings of interest. The following orthogonal loading matrix shows the difference between OMT- and OT-rotation (see Table 1, left). As the mean of the cross-loadings that balance out within each block of salient loadings is zero within each block of salient loadings, the ideal OMT-rotated loading pattern is already reached so that the initial orthogonal solution is not modified by OMT-rotation. In contrast, OT-rotation minimized the negative loadings and thereby introduces a negative factor inter-correlation (Table 1, bottom). In consequence, the block-wise mean cross-loadings of the OT-rotated solution is not zero. It is, of course, a matter of theoretical preference, which model should be used. However, it is clear that the OMT-rotated solution could also be of interest when the mean non-salient loadings are expected to be zero.

Table 1. Rotation example with initial orthogonal loadings

| variables | initial orthogonal loadings | | | OT-rotated loadings | | | OMT-rotated loadings | | |
|---|---|---|---|---|---|---|---|---|---|
| | F1 | F2 | F3 | F1 | F2 | F3 | F1 | F2 | F3 |
| $x_1$ | **.50** | .20 | -.20 | **.52** | .25 | -.11 | **.50** | .20 | -.20 |
| $x_2$ | **.50** | -.20 | .20 | **.52** | -.11 | .25 | **.50** | -.20 | .20 |
| $x_3$ | **.50** | .20 | -.20 | **.52** | .25 | -.11 | **.50** | .20 | -.20 |
| $x_4$ | **.50** | -.20 | .20 | **.52** | -.11 | .25 | **.50** | -.20 | .20 |
| $x_5$ | **.50** | .20 | -.20 | **.52** | .25 | -.11 | **.50** | .20 | -.20 |
| $x_6$ | **.50** | -.20 | .20 | **.52** | -.11 | .25 | **.50** | -.20 | .20 |
| $x_7$ | .20 | **.50** | -.20 | .25 | **.52** | -.11 | .20 | **.50** | -.20 |
| $x_8$ | -.20 | **.50** | .20 | -.11 | **.52** | .25 | -.20 | **.50** | .20 |
| $x_9$ | .20 | **.50** | -.20 | .25 | **.52** | -.11 | .20 | **.50** | -.20 |
| $x_{10}$ | -.20 | **.50** | .20 | -.11 | **.52** | .25 | -.20 | **.50** | .20 |
| $x_{11}$ | .20 | **.50** | -.20 | .25 | **.52** | -.11 | .20 | **.50** | -.20 |
| $x_{12}$ | -.20 | **.50** | .20 | -.11 | **.52** | .25 | -.20 | **.50** | .20 |
| $x_{13}$ | .20 | -.20 | **.50** | .25 | -.11 | **.52** | .20 | -.20 | **.50** |
| $x_{14}$ | -.20 | .20 | **.50** | -.11 | .25 | **.52** | -.20 | .20 | **.50** |
| $x_{15}$ | .20 | -.20 | **.50** | .25 | -.11 | **.52** | .20 | -.20 | **.50** |
| $x_{16}$ | -.20 | .20 | **.50** | -.11 | .25 | **.52** | -.20 | .20 | **.50** |
| $x_{17}$ | .20 | -.20 | **.50** | .25 | -.11 | **.52** | .20 | -.20 | **.50** |
| $x_{18}$ | -.20 | .20 | **.50** | -.11 | .25 | **.52** | -.20 | .20 | **.50** |
| | | | | | | | factor inter-correlations | | |
| F1 | 1.00 | | | 1.00 | | | 1.00 | | |
| F2 | .00 | 1.00 | | -.22 | 1.00 | | .00 | 1.00 | |
| F3 | .00 | .00 | 1.00 | -.22 | -.22 | 1.00 | .00 | .00 | 1.00 |

*Note.* Salient loadings are given in bold face.

## Simulation study

*Specification*

As OMT-rotation minimizes the weighted mean cross-loadings for each block of salient loadings, departures of single cross-loadings from the respective weighted mean cross-loadings should not affect the results as long as they are symmetrically distributed around zero. In order to investigate a condition where departures from the mean cross-loadings are due to sampling error oblique ICM-population models were investigated. It was expected that the effect of sampling error on the weighted mean cross-loadings computed in OMT-rotation is smaller than on the single cross-loadings used in OT-rotation. If these assumptions are correct, OMT-rotation should recover a population factor inter-correlations more exactly than OT.

Therefore, a simulation study based on the population ICM with $q \in \{3, 6, 9, 12\}$ factors and $p/q \in \{5, 8\}$ salient loadings per factor was performed. For $p/q = 5$ two levels of salient loadings were introduced with

$$\lambda_{.50} = \begin{bmatrix} .40 \\ .45 \\ .50 \\ .55 \\ .60 \end{bmatrix} \text{ and } \lambda_{.70} = \begin{bmatrix} .60 \\ .65 \\ .70 \\ .75 \\ .80 \end{bmatrix} \tag{10}$$

for each salient loading block with $p/q = 5$. For $p/q = 8$ the two levels of salient loadings were

$$\lambda_{.50} = \begin{bmatrix} .38 \\ .42 \\ .45 \\ .48 \\ .52 \\ .55 \\ .58 \\ .62 \end{bmatrix} \text{ and } \lambda_{.70} = \begin{bmatrix} .58 \\ .62 \\ .65 \\ .68 \\ .72 \\ .75 \\ .78 \\ .82 \end{bmatrix}. \tag{11}$$

The standard deviation of the loadings was about .08 for both levels of $p/q$. Three levels of $\phi \in \{.00, .25, .50\}$ factor inter-correlations and five sample sizes $n \in \{100, 150, 200, 300, 500\}$ were investigated. This results in 4 ($q$) × 2 ($p/q$) × 2 ($\lambda_{50}$, $\lambda_{70}$) × 3 ($\phi$) × 5 ($n$) = 240 conditions of the simulation study. The dependent variables were the OT- and OMT-factor inter-correlations which were compared with the population factor inter-correlations and the root mean square (RMS) difference of the OT- and OMT-rotated factor pattern with the population loading pattern.

Data generation was performed with the R-package 'fungible' provided by Waller (2023) based on Waller (2016), where population loadings and factor inter-correlations were entered in order to generate sample correlation matrices. For each of the 240 conditions 1,000 sample correlation matrices were generated. Least squares factor analysis with the correct number of factors was performed with 'fungible' and unrotated factor loadings were computed. The unrotated factor loadings were entered in the script as it can be found in Appendix B to compute the OT- and OMT-rotated loadings and the corresponding factor inter-correlations.

*Results*

The results for the factor inter-correlations for population ICM based on population factor inter-correlations of $\phi = .50$ are presented in Figure 1. For mean salient loadings of .50 and samples of $n = 200$ and below, the mean inter-correlations of OT-rotated factors are considerably smaller than $\phi = .50$. In contrast, the mean inter-correlations of the OMT-rotated factors are much closer to .50 and show a smaller negative bias. For q = 12 factors and mean salient loadings of .50, the mean inter-correlations of the OT-rotated factors are zero, whereas the mean inter-correlations of the OMT-rotated factors are a bit larger than .20. Thus, the under-estimation of the inter-correlations is present in all target-rotated factors but it is much smaller for the OMT-rotated factors than for the OT-rotated factors. The under-estimation of the population factor inter-correlations is considerably reduced for mean salient loadings of .70 (see Figure 1). The under-estimation of factor inter-correlations was also smaller for OMT-rotated factors than for OT-rotated factor for $\phi = .25$ (see Appendix A, Figure A1). Overall, the size of the effects was reduced and the pattern was the same as for $\phi = .50$. No under-estimation of the population factor inter-correlations and no substantial difference between OT- and OMT-rotated factors occurred for $\phi = .00$ (see Supplement, Figure S1). However, in this condition, the standard deviation of the factor inter-correlations was larger for OT-rotated factors than for OMT-rotated factors for mean salient loadings of .50, $n = 100$, and $q = 12$.

The mean RMS differences of the OT- and OMT-rotated loading patterns with the population loadings for the $\phi = .50$ condition are presented in Figure 2. For all loading sizes and sample sizes, the mean RMS differences were nearly the same for $q = 3$. For q > 6, mean salient loadings of .50, and sample sizes smaller than 300, the mean RMS differences were substantially larger for OT-rotated factor patterns than for OMT-rotated factor patterns. In these conditions, the mean RMS differences increased with $q$ for the OT-rotated factor patterns, whereas they did not substantially increase with $q$ for the OMT-rotated factor patterns. In these conditions, the standard deviations of the RMS differences were much larger for the OT-rotated factor patterns than for the OT-rotated factor patterns (see Figure 2). For $\phi = .25$ the effects of $q$, $n$, and mean salient loading size on mean RMS differences were smaller than for $\phi = .50$, but the pattern of results was the same (see Appendix A, Figure A2). For $\phi$ = .00 the mean RMS differences were very small and only a small increase of mean RMS differences occurred for OT-rotated factors for $n$ = 100, $q$ > 6, and mean salient loadings of .50.

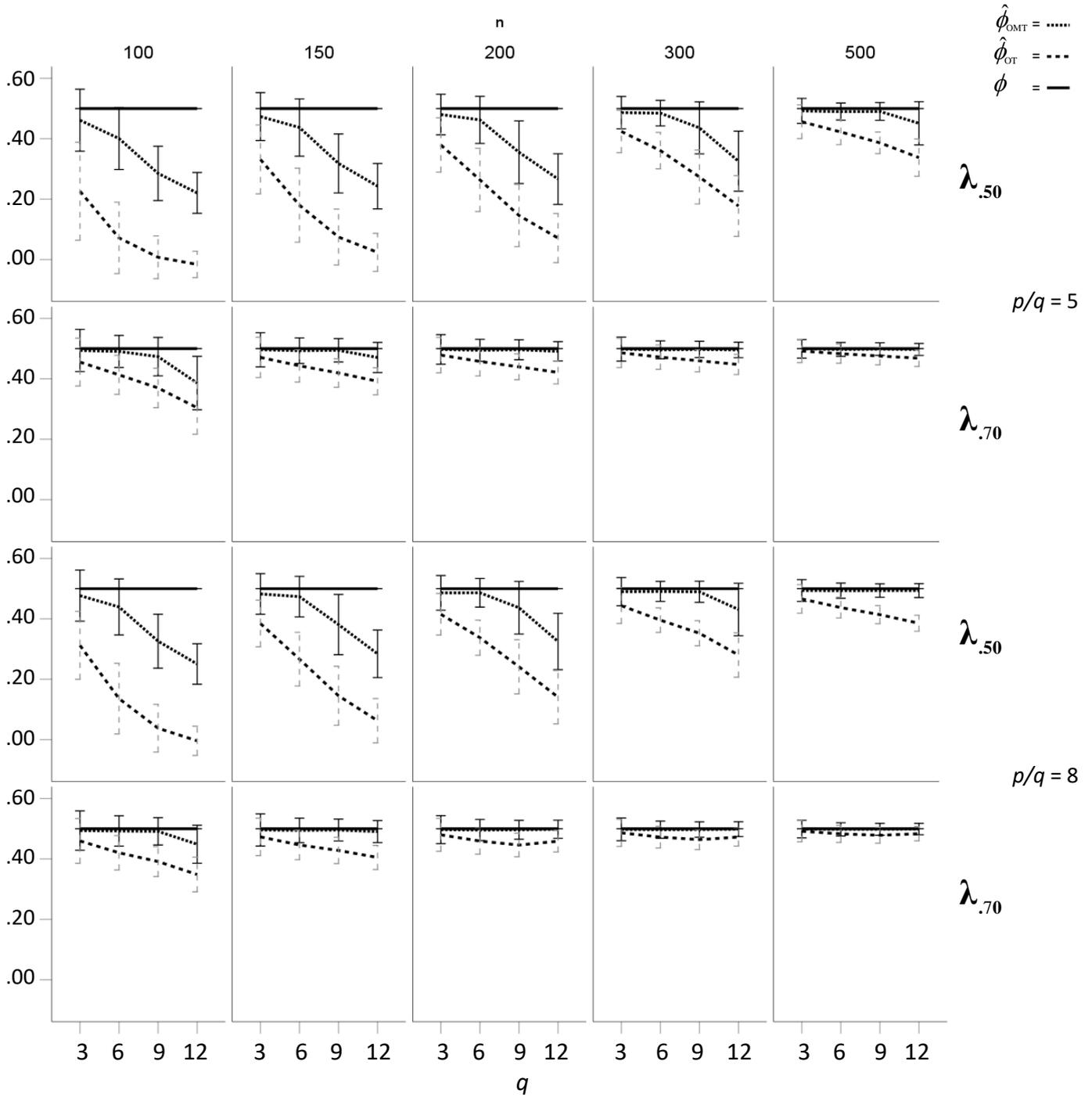

Figure 1. Means and standard deviations of inter-factor correlations resulting from OT- and OMT-rotation for population factor inter-correlations of $\phi$ = .50

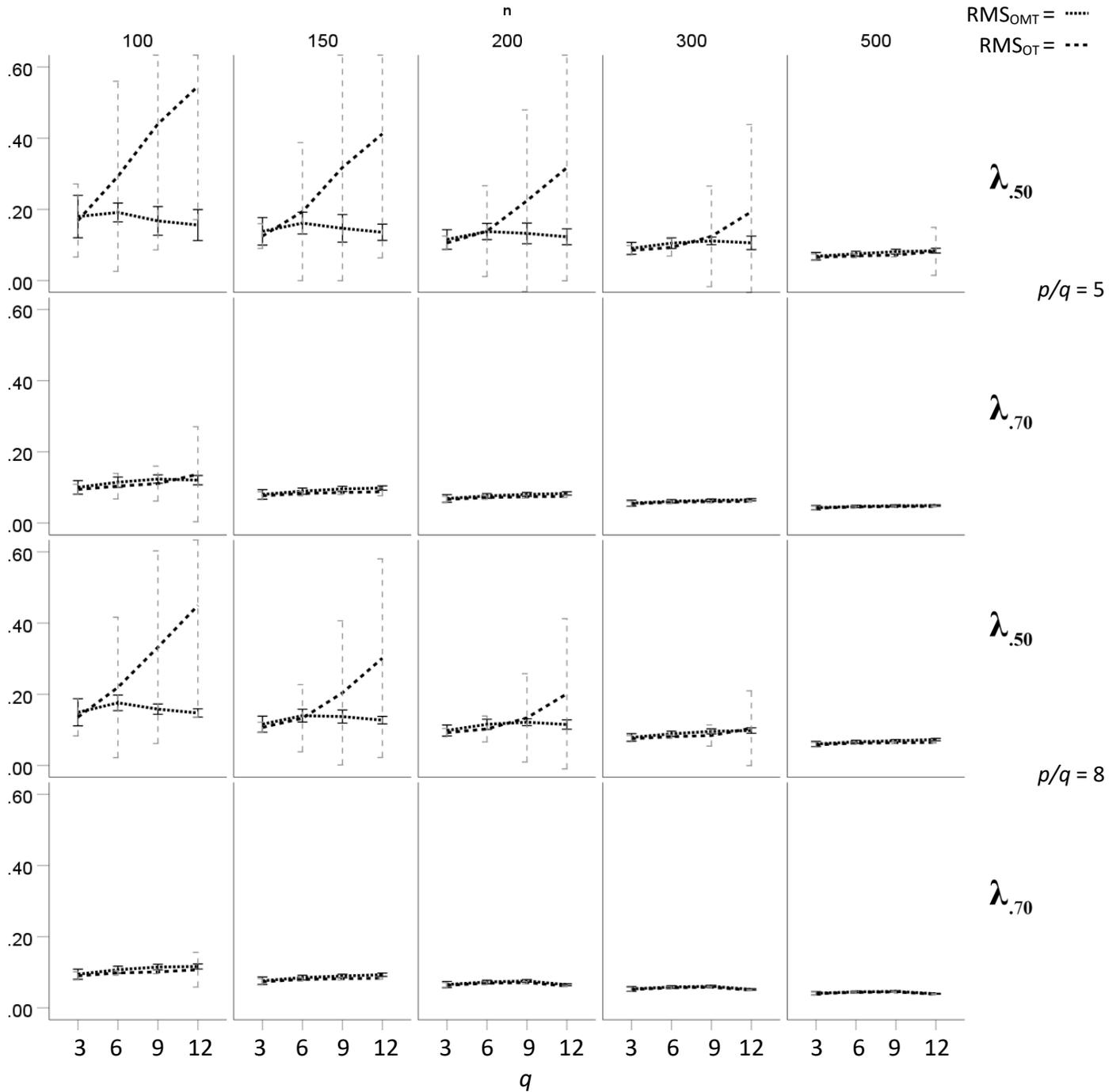

Figure 2. Root Mean Square (RMS) difference between the population loading pattern and the OT- and OMT-rotated loading patterns for population factor inter-correlations of $\phi = .50$

**Empirical Example**

As an empirical example a subsample of participants responses to 25 items from the Open-Source Psychometrics Project (http://openpsychometrics.org/_rawdata/) based on the Big-Five Factor

Markers (BIG5.zip, last updated 5/18/2014, retrieved on 08/22/2023) from the International Personality Item Pool (IPIP, Goldberg, 1992) was used. Only the first 19,700 participants (age/years: $M = 26.27$, $SD = 11.59$; gender: 11,973 females, 7,601 males, 102 others, 24 missing values) from the total file of 19,719 participants were used in order to split the total sample into 197 subsamples each containing the responses of 100 participants to the first four items (E1-E4, N1-N4, A1-A4, C1-C4, O1-O4) of each of the five factors. Only a subsample of items was used in order to investigation a data set that is less favorable for optimal factor rotation.

The aim was to compare the OT- and OMT-rotated five-factor solution of the total sample with the OT- and OMT-rotated five-factor solutions of the subsamples. Principal axis factoring of the total sample and of the subsamples was performed with IBM SPSS Version 29.0 and OT- and OMT-rotation was performed with the code provided in Appendix B. The rotated solutions for the total sample are presented in Table 2. The OT- and OMT-rotated loading patterns are very similar which indicates that for the very large total sample both rotation methods work well. The inter-correlations of the OMT-rotated factors were a bit larger than the inter-correlations of the OT-rotated factors.

Overall, 195 out of the 197 principal factor analyses converged. OT- and OMT-rotation was performed for the unrotated factor solutions and the RMS difference of each of the rotated factor patterns with the corresponding rotated factor pattern of the total sample was computed. When for $RMS_{OT}$ five values greater one were set to one, the mean of $RMS_{OT}$ was 0.18 ($SD = 0.19$), for $RMT_{OT}$ no values greater one occurred and the mean $RMS_{OMT}$ was .16 ($SD = .06$). For the factor inter-correlations of the OT-rotated factors $RMS_{OT}$ was 0.25 ($SD = 0.21$, two values greater one were set to one), for the factor inter-correlations of the OMT-rotated factors $RMS_{OMT}$ was 0.15 ($SD = 0.05$, no values greater one occurred).

Table 2. OT- and OMT-rotated five factor loading patterns and factor inter-correlations 6for 20 BIG-Five Markers of the total sample

|    | OMT-rotation | | | | | OT-rotation | | | | |
|----|------|------|------|------|------|------|------|------|------|------|
|    | E    | N    | A    | C    | O    | E    | N    | A    | C    | O    |
| E1 | **.71** | .03 | -.12 | -.06 | -.01 | **.66** | -.03 | -.02 | -.01 | .01 |
| E2 | **.75** | .15 | -.10 | -.07 | .03 | **.69** | .09 | .01 | -.01 | .05 |
| E3 | **.65** | -.13 | .14 | .06 | -.06 | **.62** | -.19 | .23 | .09 | -.04 |
| E4 | **.77** | .04 | -.15 | .03 | .02 | **.72** | -.02 | -.05 | .08 | .04 |
| N1 | .02 | **.76** | -.01 | -.02 | -.04 | .00 | **.75** | -.01 | -.02 | -.06 |
| N2 | -.02 | **.60** | -.10 | .08 | .00 | -.03 | **.60** | -.11 | .09 | -.01 |
| N3 | -.02 | **.72** | .09 | .04 | .00 | -.03 | **.72** | .08 | .03 | -.01 |
| N4 | -.09 | **.34** | -.01 | -.05 | .12 | -.10 | **.34** | -.02 | -.04 | .11 |
| A1 | .03 | .05 | **.39** | .00 | .11 | .03 | .04 | **.40** | .00 | .12 |
| A2 | .34 | .03 | **.43** | -.08 | .05 | .32 | .00 | **.48** | -.07 | .06 |
| A3 | -.17 | -.16 | **.45** | .22 | -.07 | -.14 | -.15 | **.42** | .19 | -.06 |
| A4 | -.02 | .07 | **.80** | -.07 | -.04 | -.02 | .05 | **.79** | -.10 | -.03 |
| C1 | .06 | .04 | -.01 | **.53** | .04 | .09 | .04 | .00 | **.54** | .05 |
| C2 | -.05 | .02 | -.06 | **.58** | -.11 | -.01 | .04 | -.07 | **.57** | -.10 |
| C3 | -.01 | .12 | .05 | **.36** | .18 | .01 | .13 | .06 | **.37** | .18 |
| C4 | .04 | -.16 | .02 | **.66** | -.04 | .08 | -.15 | .03 | **.66** | -.02 |
| O1 | .06 | .05 | -.06 | .03 | **.47** | .06 | .04 | -.03 | .08 | **.46** |
| O2 | -.01 | -.08 | -.05 | .05 | **.74** | .00 | -.08 | -.01 | .11 | **.74** |
| O3 | .04 | .10 | .01 | -.11 | **.38** | .03 | .09 | .03 | -.07 | **.38** |
| O4 | -.04 | -.02 | .06 | -.04 | **.64** | -.03 | -.02 | .08 | .00 | **.63** |
|    | OMT factor inter-correlations | | | | | OT factor inter-correlations | | | | |
| E  | 1.00 | | | | | 1.00 | | | | |
| N  | -.31 | 1.00 | | | | -.20 | 1.00 | | | |
| A  | .35 | .00 | 1.00 | | | .20 | .03 | 1.00 | | |
| C  | .15 | -.24 | .18 | 1.00 | | -.02 | -.25 | .18 | 1.00 | |
| O  | .11 | -.14 | .16 | .11 | 1.00 | .04 | -.09 | .08 | -.01 | 1.00 |

## Discussion

Investigations of the ICM by means of EFA are still relevant, also because analyses of the ICM by means of confirmatory factor analyses may lead to series of model-modifications. It was, however, expected that sampling error substantially affects results of conventional Target-rotation because single cross-loadings are minimized. In order to reduce the effect of sampling error on results, OMT-rotation was proposed which minimizes mean cross-loadings instead of single cross-loadings. It was shown in a population example that minimizing single cross-loadings by means of conventional OT-rotation may lead to ambiguous results, when the mean cross-loadings are close to zero while the absolute size of the cross-loadings is substantial. In the population model, the observed variables with single cross-loadings that were close to zero after rotation were arbitrary because the variables all had the same absolute cross-loading before rotation. This indicates that OMT-rotation may be of special interest when the cross-loadings with positive and negative sign balance out.

The simulation study for oblique ICM reveals that sampling error may induce negative bias to the Target-rotated factor inter-correlations. The negative bias of the factor inter-correlations was substantially more pronounced for OT-rotation than for OMT-rotation, especially for small sample sizes, moderate mean salient loadings, and a large number of factors. For 12 factors, 100 cases, mean salient loadings of .50, and population inter-correlations of .50, the mean sample inter-correlations of OT-rotated factors was zero, whereas it was greater .20 for OMT-rotated factors. The mean RMS differences of rotated factor patterns and the population factor pattern were larger for OT-rotation than for OMT-rotation. Thus, when samples size was small and the number of factors large, loading patterns and factor inter-correlations were more similar to the population loading patterns and factor inter-correlations for OMT-rotation than for OT-rotation. However, no relevant differences between the rotation methods were found for the uncorrelated ICM.

An empirical example was based on open data for the BIG-five model of personality (Goldberg, 1992). A large total sample based on four marker variables per factor was divided into several subsamples based on 100 participants in order to investigate the similarity of the OT- and OMT-rotated subsample solutions with the corresponding OT- and OMT-rotated total sample solutions. The similarity of the rotated loading patterns and factor inter-correlations for the subsamples with the corresponding rotated loading pattern and factor inter-correlations for the total sample was more pronounced for OMT-rotation than for OT-rotation. This indicates that OMT-rotation may help to get more robust results, especially when the number of marker variables per factor and the sample size are rather small.

Overall, the results of the simulation study and of the empirical example indicate that OMT-rotation is more robust than OT-rotation. Therefore, OMT-rotation can be recommended when an oblique ICM is expected, when salient loadings are moderate, factor numbers large, and sample sizes small. The relevant orthogonal/unrotated loading matrices for OMT-rotation may be entered into the R-script or into the SPSS-script provided in Appendix B.

A limitation of the present study is that it is restricted to the ICM. It might be of interest to compare OMT- and OT-rotation for other population factor models. It should, however, be noted that more complex factor models may not allow to draw clear conclusions on optimal factor inter-correlations. The reason is that for complex models some researchers might prefer larger cross-loadings and smaller factor loadings and others might prefer smaller cross-loadings and larger factor inter-correlations (Schmitt and Sass 2011). There is typically no objective way to decide between these preferences. It should, however, be noted that it is possible to exclude cross-loadings from the computation of the mean non-salient loadings in OMT-rotation. This would make sense when a substantial cross-loading is in line with the theoretical expectations so that it should not be reduced by means of factor rotation. Thereby, partial OMT-rotation could be performed.

From a broader perspective, it should be noted that less biased factor inter-correlations are an important basis for hierarchical factor models. Factor prediction, as it can be performed with ESEM, also needs optimal estimates of the factor inter-correlations. Finally, correlation-preserving factor score predictors (e.g., McDonald 1981) also require optimal estimates of the factor inter-correlations. Especially in settings where the factor inter-correlations are relevant for further research OMT-rotation might be considered.

# Appendix A

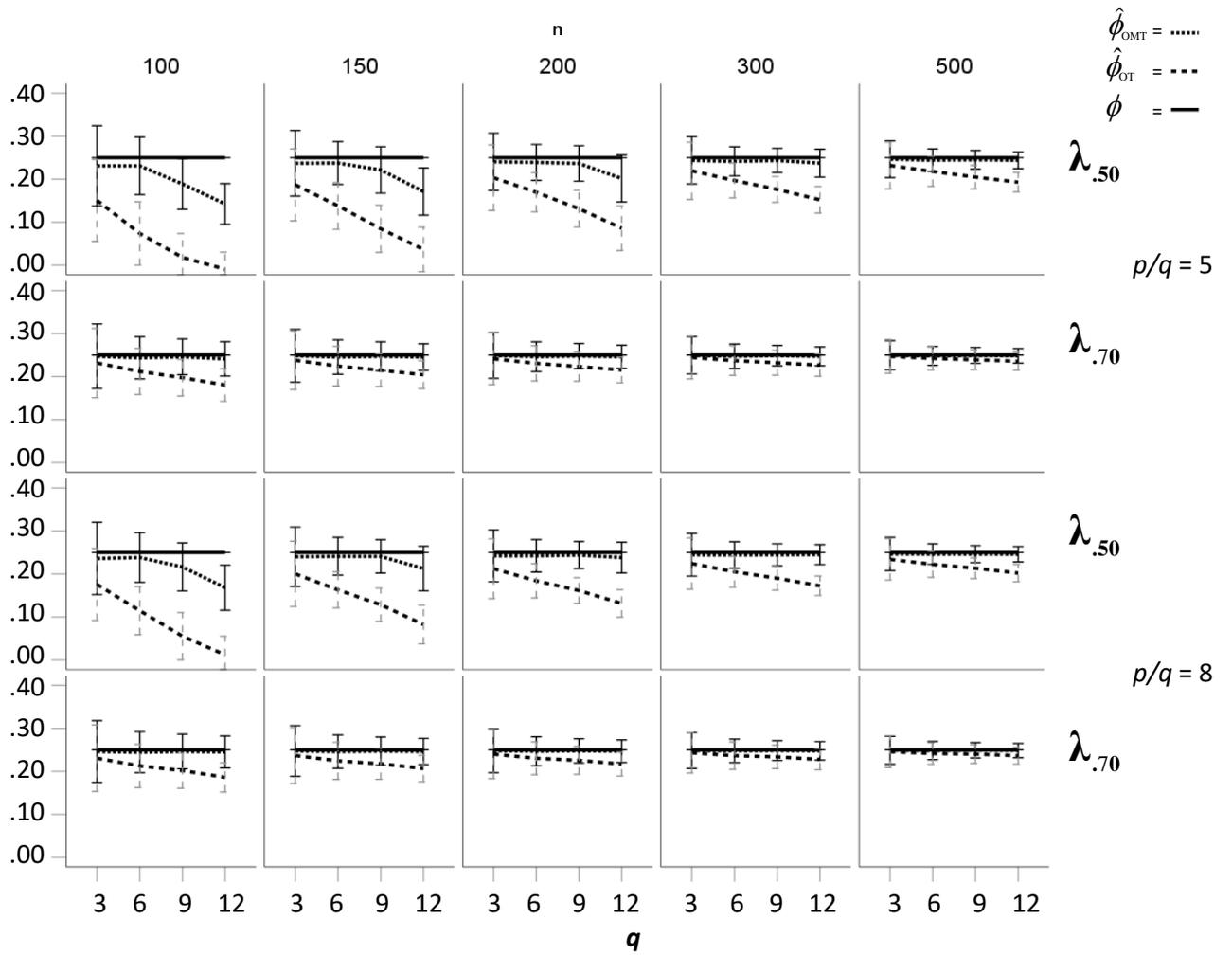

Figure A1. Means and standard deviations of inter-factor correlations resulting from OT- and OMT-rotation for population factor inter-correlations of $\phi = .25$

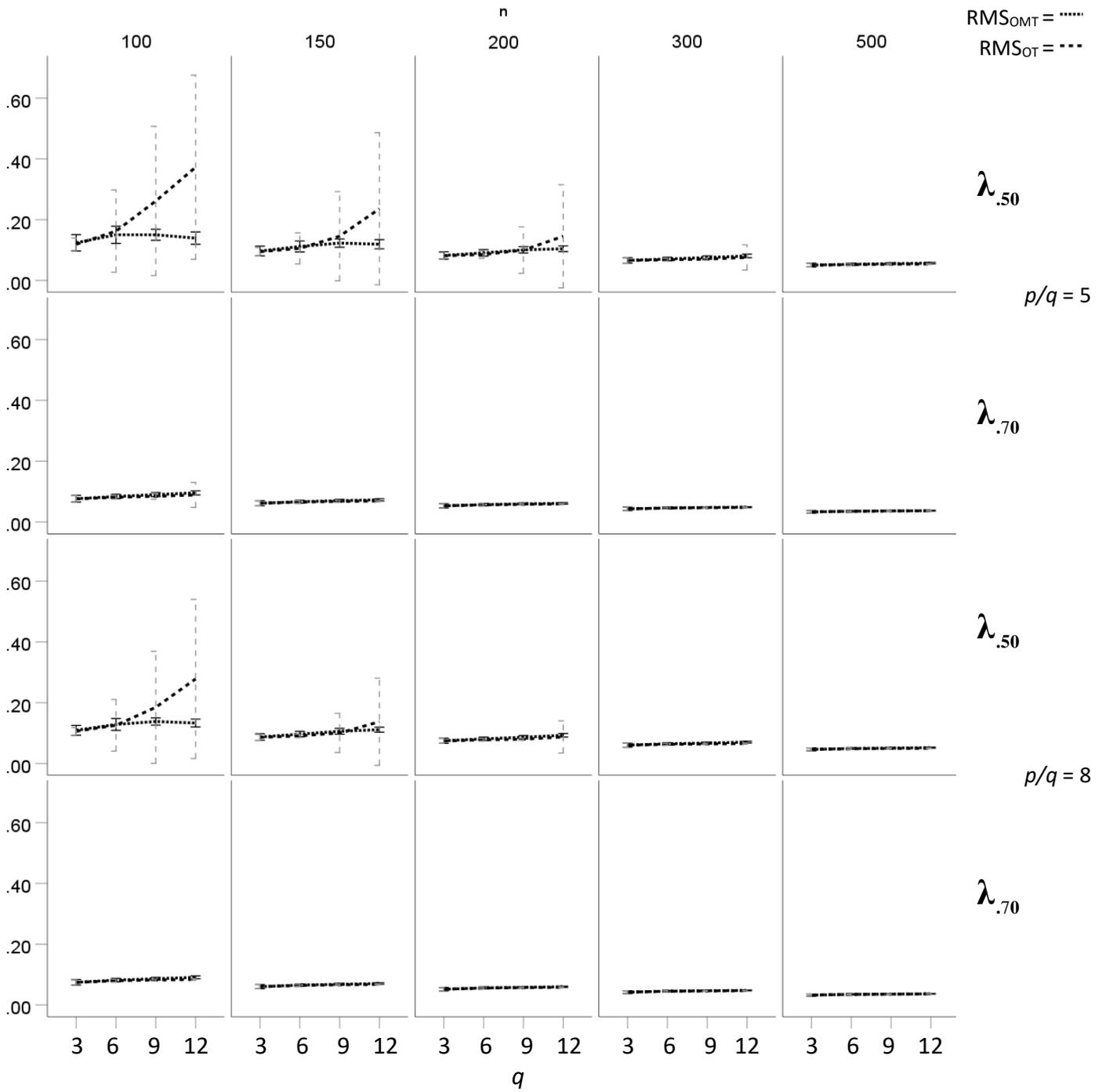

Figure A2. Root Mean Square (RMS) difference between the population loading pattern and the OT- and OMT-rotated loading patterns for population factor inter-correlations of $\phi = .25$

# Appendix B

**# R-Script:**

```
# Context: R-4.3.1, the following packages are needed:

library(ramify)
library(RSpectra)
library(matrixcalc)
library(fastmatrix)

# Helper functions for frequently used matrix operations
Mdiag <- function(x) return(diag(diag(x)))
inv <- function(x) return(solve(x))
helpPhi <- function(x) return( inv(t(x)%*%x)%*%t(x)%*% L%*%t(L)
      %*%x%*%inv(t(x)%*%x) )

# Enter values for Oblique Mean Target (OMT)-Rotation:

# number of factors:
q <- 3

# number of variables:
p <- 18

# Enter repetitions for Ridge-constant (reducing Kappa):
Iterate <- 100

# Enter Kappa that might be regarded as too large (default = 20):
k_level <- 20

# Enter inital orthogonal loadings L_u for Target-rotation.

L <- matrix(0, nrow= p, ncol= q)
L[1,]  <- c( 0.50, 0.20,-0.20)
L[2,]  <- c( 0.50,-0.20, 0.20)
L[3,]  <- c( 0.50, 0.20,-0.20)
L[4,]  <- c( 0.50,-0.20, 0.20)
L[5,]  <- c( 0.50, 0.20,-0.20)
L[6,]  <- c( 0.50,-0.20, 0.20)
L[7,]  <- c( 0.20, 0.50,-0.20)
L[8,]  <- c(-0.20, 0.50, 0.20)
L[9,]  <- c( 0.20, 0.50,-0.20)
L[10,] <- c(-0.20, 0.50, 0.20)
L[11,] <- c( 0.20, 0.50,-0.20)
L[12,] <- c(-0.20, 0.50, 0.20)
L[13,] <- c( 0.20,-0.20, 0.50)
L[14,] <- c(-0.20, 0.20, 0.50)
L[15,] <- c( 0.20,-0.20, 0.50)
L[16,] <- c(-0.20, 0.20, 0.50)
L[17,] <- c( 0.20,-0.20, 0.50)
L[18,] <- c(-0.20, 0.20, 0.50)
print(round(L,2))

# Orthogonal factors:
Phi <- diag(1,q)

# Enter ICM-Target-matrix: "1" for salient-loading, "0" for non-salient
      loadings.
```

```
# For block-diagonal Target-matrices you may use:.
#IDmat <- diag(1,q)
#Tar <- kronecker.prod(IDmat, matrix(1,p/q,1))
# More complex Target-matrices can be entered directly:.
Tar <- matrix(0, nrow= p, ncol= q)
Tar[1,]  <- c(1, 0, 0)
Tar[2,]  <- c(1, 0, 0)
Tar[3,]  <- c(1, 0, 0)
Tar[4,]  <- c(1, 0, 0)
Tar[5,]  <- c(1, 0, 0)
Tar[6,]  <- c(1, 0, 0)
Tar[7,]  <- c(0, 1, 0)
Tar[8,]  <- c(0, 1, 0)
Tar[9,]  <- c(0, 1, 0)
Tar[10,] <- c(0, 1, 0)
Tar[11,] <- c(0, 1, 0)
Tar[12,] <- c(0, 1, 0)
Tar[13,] <- c(0, 0, 1)
Tar[14,] <- c(0, 0, 1)
Tar[15,] <- c(0, 0, 1)
Tar[16,] <- c(0, 0, 1)
Tar[17,] <- c(0, 0, 1)
Tar[18,] <- c(0, 0, 1)
#print(round(Tar,2))

# Oblique Mean Target (OMT)-Rotation:
#1 orthogonal Target-rotation, Schönemann, 1966:

S <- t(L)%*%(Tar)
help1e <- eigen(S%*%t(S))
WW <- help1e$vectors
help2 <- t(S)%*%S
help2e <- eigen(help2)
V <- help2e$vectors
O <- t(WW)%*%S%*%V
ON <- ((O)/abs(O+0.00000000000001))
K <- diag(diag(ON))
WWW <- (WW%*%K)
TR <- WWW%*%t(V)
L1 <- L%*%TR

#2 weigthed average loadings per ICM-cluster according to Equation 4:
L1m <- t(L1 * Tar) %*% (L1) %*% inv(t(L1 * Tar) %*% Tar)
L1m
# Target-matrix for oblique Target-rotation
Iq <- diag(1,q)

#3 oblique rotation of averaged loadings, Target-rotation, Hurley & Cattell,
     1962:
# Equation 5:
cong_OMT_old <- 0
help <- t(L1m)%*%L1m
for (ii in 1:Iterate) {
if (kappa(help) > k_level) {
help <- help + diag(0.01,q)
}
T <- inv(help)%*%t(L1m)%*%Iq
# normalize transformation matrix, Equation 6:
Tn <- inv(Mdiag(t(T)%*%T)^0.5)%*%T
# reference structure, Equation 7:
L2 <- L1%*%Tn
# OMT- loading pattern according to Equation 8:
```

```
L_OMT <- L2 %*% Mdiag(inv(t(Tn)%*%Tn))^0.5
# Equation 9:
cong_OMT <- matrix.trace( t(L_OMT)%*%Tar  %*%inv( Mdiag(t(L_OMT)%*%L_OMT) %*%
      Mdiag(t(Tar)%*%Tar) )^0.5) / q
if (cong_OMT > cong_OMT_old) {
L_OMT_old <- L_OMT
cong_OMT_old <- cong_OMT
}
}
cong_OMT <- cong_OMT_old
L_OMT <- L_OMT_old
Phi_OMT <- helpPhi(L_OMT)
round(L_OMT,2)
round(Phi_OMT,2)
round(cong_OMT, 3)

# Oblique Target-rotation (OT), Hurley & Cattell, 1962:
TT <- inv(t(L)%*%L)%*%t(L)%*%Tar
check <- t(TT)%*%TT
D <- Mdiag(check)^0.5
# normalize transformation matrix
TT <- inv(D)%*%TT
# reference structure
FTT <- L%*%TT
CR <- t(TT)%*%TT
D <- inv(Mdiag(inv(CR))^0.5)
# factor pattern and phi
L_OT <- FTT%*%inv(D)
Phi_OT <- helpPhi(L_OT)
round(L_OT,2)
round(Phi_OT,2)
cong_OT <- matrix.trace( t(L_OT)%*%Tar  %*%inv( Mdiag(t(L_OT)%*%L_OT) %*%
      Mdiag(t(Tar)%*%Tar) )^0.5) / q
round(cong_OT, 3)
```

**\* SPSS-Script:**

```
* Encoding: windows-1252.

SET MXLOOPS = 100.
MATRIX.

/* Enter values for Oblique Mean Target (OMT)-Rotation .
/* Enter repetitions for Ridge-constant (reducing kappa) .
compute Iterate = 100.
/* Enter kappa that might be regarded as too large (default = 20) .
compute k_level = 20.
/* Enter inital orthogonal loadings L_u for Target-rotation .
compute L_u = {.50,  .20, -.20;
               .50, -.20,  .20;
               .50,  .20, -.20;
               .50, -.20,  .20;
               .50,  .20, -.20;
               .50, -.20,  .20;
               .20,  .50, -.20;
              -.20,  .50,  .20;
               .20,  .50, -.20;
              -.20,  .50,  .20;
               .20,  .50, -.20;
              -.20,  .50,  .20;
               .20, -.20,  .50;
              -.20,  .20,  .50;
               .20, -.20,  .50;
              -.20,  .20,  .50;
               .20, -.20,  .50;
              -.20,  .20,  .50}.
compute p = nrow(L_u).
compute q = ncol(L_u).
/* Enter ICM-Target-matrix: "1" for salient-loading, "0" for non-salient
      loadings .
/* For block-diagonal Target-matrices you may use:.
/* compute Tar = kroneker(ident(q),make(p/q,1,1)).
/* More complex Target-matrices can be entered directly:.
compute Tar = {1, 0, 0;
               1, 0, 0;
               1, 0, 0;
               1, 0, 0;
               1, 0, 0;
               1, 0, 0;
               0, 1, 0;
               0, 1, 0;
               0, 1, 0;
               0, 1, 0;
               0, 1, 0;
               0, 1, 0;
               0, 0, 1;
               0, 0, 1;
               0, 0, 1;
               0, 0, 1;
               0, 0, 1;
               0, 0, 1
}.
/* Orthogonal factors .
compute phi = ident(q).

/* Oblique Mean Target (OMT)-Rotation .
/* [1] Orthogonal Target-rotation (Schönemann, 1966) .
compute S = t(L_u)*Tar.
```

```
call eigen(S*t(S),WW,lamb).
call eigen(t(S)*S,V,lamb).
compute O = t(WW)*S*V.
compute ON = ((O)/abs(O+1E-14)).
compute K = mdiag(diag(ON)).
compute WWW = WW*K.
compute TR = WWW*t(V).
compute L1 = L_u*TR.

/* [2] Weigthed average loadings per ICM-cluster (Equation 4) .
compute L1m = t(L1&*Tar)*(L1)*inv(t(L1&*Tar)*Tar).
/* Target-matrix for oblique Target-rotation .
compute Iq = mdiag(make(q,1,1)).

/* [3] Oblique rotation of averaged loadings, Target-rotation (Hurley &
      Cattell, 1962) .
/* Equation 5 .
compute cong_OMT_old = 0.
compute help = t(L1m)*L1m.
loop ii = 1 to Iterate.
   compute T_ = inv(help)*t(L1m)*Iq.
/* Normalize transformation matrix (Equation 6) .
   compute Tn = inv(mdiag(diag(t(T_)*T_))&**0.5)*T_.
/* Reference structure (Equation 7) .
   compute L2 = L1*Tn.
/* OMT-loading pattern (Equation 8) .
   compute L_OMT = L2*mdiag(diag(inv(t(Tn)*Tn)))&**0.5.
/* Equation 9 .
   compute cong_OMT =
      trace(t(L_OMT)*Tar&*inv(mdiag(diag(t(L_OMT)*L_OMT))&*mdiag(diag(t(Tar)*
      Tar)))&**0.5)/q.
   call eigen(help,vec,val).
   compute kappa = abs(mmax(val))/abs(mmin(val)).
   do if (cong_OMT > cong_OMT_old).
      compute L_OMT_old = L_OMT.
      compute cong_OMT_old = cong_OMT.
   end if.
   do if kappa > k_level.
      compute help = help+mdiag(make(q,1,0.01)).
   else.
      break.
   end if.
end loop.
compute cong_OMT = cong_OMT_old.
compute L_OMT = L_OMT_old.
compute Phi_OMT =
      inv(t(L_OMT)*L_OMT)*t(L_OMT)*L_u*t(L_u)*L_OMT*inv(t(L_OMT)*L_OMT).

/* Oblique Target-rotation (OT) (Hurley & Cattell, 1962) .
compute TT = inv(t(L_u)*L_u)*t(L_u)*Tar.
compute check = t(TT)*TT.
compute D = mdiag(diag(check))&**0.5.

/* Normalize transformation matrix .
compute TT = inv(D)*TT.

/* Reference structure .
compute FTT = L_u*TT.
compute CR = t(TT)*TT.
compute D = inv(mdiag(diag(inv(CR))))&**0.5.

/* Factor pattern and phi .
compute L_OT = FTT*inv(D).
```

```
compute Phi_OT = inv(t(L_OT)*L_OT)*t(L_OT)*L_u*t(L_u)*L_OT*inv(t(L_OT)*L_OT).
compute cong_OT =
      trace(t(L_OT)*Tar&*inv(mdiag(diag(t(L_OT)*L_OT))&*mdiag(diag(t(Tar)*Tar
      )))&**0.5)/q.

print L_u /formats=f5.2.
print L1m /formats=f5.2.
print L_OMT /formats=f5.2.
print Phi_OMT /formats=f5.2.
print cong_OMT /formats=f5.2.
print L_OT /formats=f5.2.
print Phi_OT /formats=f5.2.
print cong_OT /formats=f5.2.

END MATRIX.
```